\documentclass{aa}
\usepackage{graphics}

\begin{document}

   \thesaurus{06     
              (02.13.3;  
               03.20.2;  
               05.01.1; 
               08.03.4;  
               08.09.2 U~Her;  
               08.16.4)}  
   \title{VLBI Astrometry of the Stellar Image of U~Herculis,
               Amplified by the 1667 OH Maser}
   \titlerunning{VLBI Astrometry of the Amplified Stellar Image of U~Her}


   \author{H.J. van Langevelde\inst{1}\and
        W. Vlemmings\inst{2}\and
        P.J. Diamond\inst{3}\and
        A. Baudry\inst{4}\and
        A.J. Beasley\inst{5}
          }

   \offprints{WV (vlemming@strw.leidenuniv.nl)}

   \institute{Joint Institute for VLBI in Europe, Postbus 2, 
                7990~AA Dwingeloo
        \and
              Sterrewacht Leiden, Postbus 9513, 2300 RA Leiden, 
              the Netherlands
        \and
        MERLIN/VLBI National Facility, Jodrell Bank Observatory, Macclesfield,
                    Cheshire, SK11 9DL, England
        \and
        Observatoire de Bordeaux, BP 89, F--33270 Floirac, France
   \and 
   National Radio Astronomy Observatory, 520 Edgemont Road, Charlottesville, VA 22903-2475, USA
        }

   \date{Received ; accepted }

   \maketitle


\begin{abstract} 
  
  The OH 1667 MHz maser in the circumstellar shell around the Mira
  variable U~Her has been observed with the NRAO Very Long Baseline
  Array (VLBA) at 6 epochs, spread over 4 years. Using phase
  referencing techniques the position of the most blue-shifted maser
  spot was monitored with respect to two extra-galactic radio sources.
  The absolute radio positions of the maser can be compared with the
  stellar optical position measured by the Hipparcos satellite to 15
  mas accuracy. This confirms the model in which one of the maser
  spots corresponds to the stellar continuum, amplified by the maser.
  The stellar proper motion and the annual parallax ($\pi_{\rm VLBI} =
  5.3 \pm 2.1$ mas) were measured.


  \end{abstract} 

\section{Introduction}
\label{intro} 

Until recently, distances to Mira variables were mostly based on the
Period -- Luminosity relation. Primary distance measurements are
important to discuss the calibration and origin of this relation,
which bears on the understanding of the structure and evolution of
stars on the AGB. Prior to the Hipparcos mission there were very few
measurements of parallaxes of Mira stars. For a number of nearby Miras
the distances are now better known; however, in several cases and for
U Her in particular, the optical parallax and hence the distance is
very uncertain.  In this paper we show that it is possible to obtain
fundamental measurements of AGB star properties by monitoring the OH
maser positions with Very Long Baseline Interferometry (VLBI).

The Mira variable U~Her is a well known source for studies of
circumstellar masers. Both its main line OH and water masers have been
the target of a number of VLBI studies (e.g. Chapman et
al. \cite{Chapman}, Yates \& Cohen \cite{YatCoh}). It has been assumed
to be relatively close; Chapman et al.\ \cite{Chapman} give a value of
385 pc based on the Mira Period -- Luminosity relation. Using a
revised P -- L relation Alvarez \& Mennesier (\cite{AlMe}) find a
value of 280 pc.

In order to use the maser positions to monitor the trajectory of
U~Her, an assumption has to be made about the motion of the masers
with respect to the underlying star. If the most blue-shifted
circumstellar maser spot corresponds to a special condition, maser
action in the shell initiated by the radio continuum radiation from
the photosphere, then this most blue-shifted spot should be a bright
beacon, necessarily fixed on the true stellar position. VLBI
observations by Sivagnanam et al.\ (\cite{SivaAmpl}) provided strong
evidence for this. They showed that in U~Her the dominant OH 1665 and
1667 MHz VLBI features at the most blue-shifted side of the spectrum
are coincident, in accordance with such a model. Observations
presented here support the amplified stellar image model, although
some questions still remain.

Even with considerable errors on the
parallax, the Hipparcos data on U~Her allow a comparison between the
optical and maser positions with unprecedented accuracy. Traditionally
the position of the star with respect to the maser features, as well
as the relative positions of maser features of different species, had
to be assumed, for instance by fitting a shell and assuming the star
is in the center. It has now become possible to overlay these
positions directly to 15 mas accuracy.

\begin{figure*} 
   \resizebox{\hsize}{!}{\includegraphics{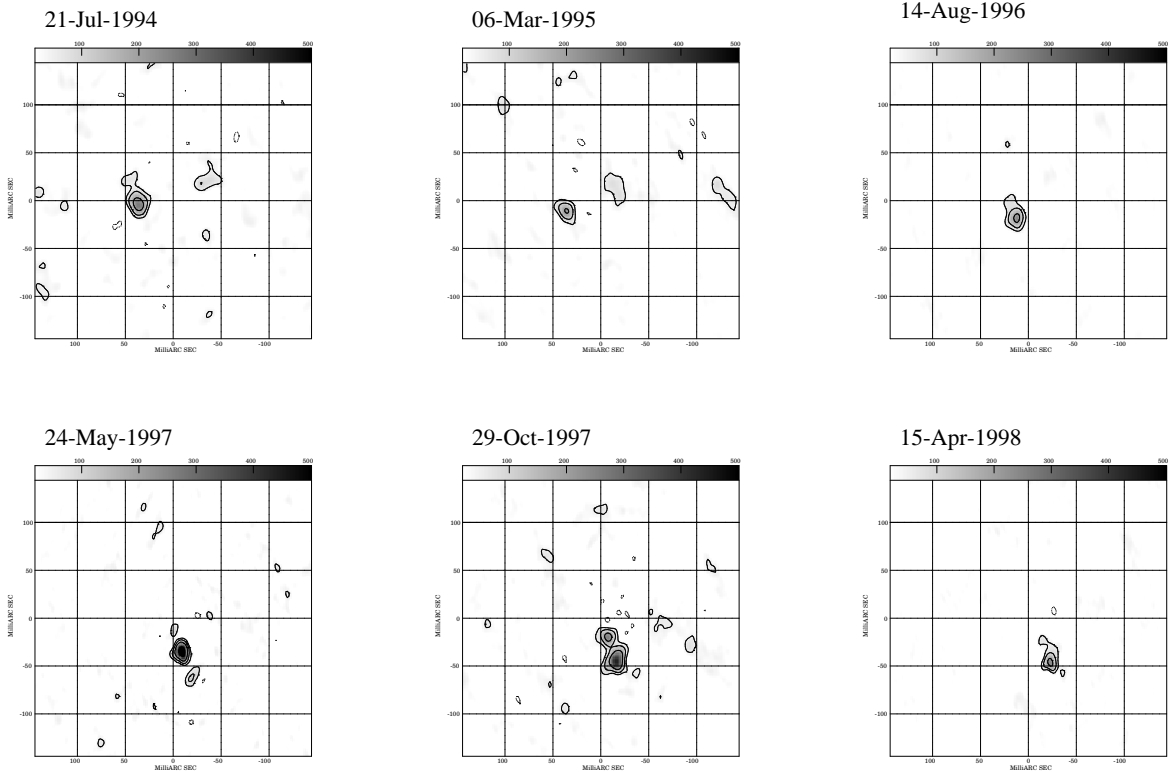}}
   \hfill
   \caption{The motion of the most blue-shifted 1667~MHz maser
   spot at $v_{\rm lsr}=-20.6$ km/s with respect to one of the
   extra-galactic reference sources. Images are 300x300 mas with a
   flux scale of 0 to 500 mJy}  \label{motion}
\end{figure*}

\section{Observations}
\label{obs}

The NRAO\footnote{The National Radio Astronomy Observatory is a
  facility of the National Science Foundation operated under
  cooperative agreement by Associated Universities, Inc.} VLBA was
  used to monitor the positions of the circumstellar masers of U~Her
  over a period of almost 4 years. In total 7 observations were
  recorded and correlated, but 6 were technically successful. These
  are July 22 1994, March 6 1995, August 8 1995, May 24 1997, October
  29 1997 and April 15 1998. For most epochs a total of 6 hours was
  used to measure U~Her. An additional 6 hours usually targeted R~LMi,
  but this source was only detected in the first epoch. In the last
  epoch R~LMi was omitted and we observed the water maser in U~Her
  instead. It was detected, but no proper phase connection with any of
  the reference sources could be made.

The 1667~MHz OH line was observed with a small bandwidth of 500 kHz
and correlated with high ($0.18$ km/s) spectral
resolution. Simultaneously three 4 MHz wide bands were recorded,
which yielded sufficient bandwidth to detect the continuum reference
sources. These were correlated separately with modest spectral
resolution (ranging from 5.6 -- 22.5 km/s). The 1665 MHz OH transition,
which is usually also excited in Mira variables exhibiting 1667 MHz maser
emission, happens to fall in one of these continuum bands.  Due to a
technical problem the first epoch is single polarization, the others
all have both circular polarizations.

Two nearby extra-galactic continuum sources were used to calibrate
phase, delay and phase rate. These were 1636+2112 and 1628+214
from Wilkinson et al.\ (\cite{Wilkinson}). The accuracy of the
calibrator positions was uncertain to $\approx$ 50 mas with respect to
fundamental VLBI reference positions (Ma et al.\ \cite{Ma}) according
to the authors. Both calibrators are within 3$\degr$ of U~Her and were
observed every 5 -- 7 minutes. Part of the schedule was done by
cycling over three sources, part by using one of the reference sources
and the target only.

Recently, the VLBA was used to determine the positions of our
calibrators with higher accuracy by phase-referencing them to
1639+230.  ($(\alpha,\delta)_{2000}=$ $16^h41^m25^s.2276$,
$+22\degr57\arcmin04\arcsec.032$).  These observations refined the
calibrator positions to be $(\alpha,\delta)_{2000}=$
$16^h36^m38^s.18373$, $+21\degr12\arcmin55\arcsec.5991$ and
$16^h30^m11^s.23117$,$+21\degr31\arcmin34\arcsec.3144$ for 1636+2112
and 1628+214 respectively. The positional accuracy is estimated to be
$\approx 2$ mas, mostly due to extended structure of the reference
sources themselves. The reference source 1639+230 was observed as part
of the USNO 9903 catalog (electronic publication). It can be tied to
the radio reference frame with $< 1$ mas accuracy; including the
measurement uncertainties, the phase reference sources can be assumed
to have {\it absolute} positions accurate up to $\approx 3$ mas. The new
positions were used to calibrate the positions determined for the
maser features.

The U~Her data have been processed in AIPS without much special astrometric
software.  We rely on the accuracy of the VLBA correlator model and
work with the residual phases directly. A special task was written to
connect the calibration of the wide band data on the reference sources
to the spectral line data sets.  In all epochs phase referencing works
well, even at this relatively low frequency; note that no ionospheric
model is included in the correlator model. The effects of the
ionosphere are seen as slightly distorted images.
Indeed, when the maser spots are bright enough, these distortions can
be corrected for by self-calibration. The resulting beam size varies
from one epoch to the other and depends on whether we can find phase
connections for the outer antennas; typically the beam is $11 \times
6$ mas.

The internal consistency of our procedures can be verified by
measuring the separation between the two calibrators at all epochs.
The difference between the measured separation and its a-priori value
was $17.7 \pm 1.6$ mas using the original positions from Wilkinson et
al. (1998). This improved to $1.8 \pm 1.6$ after new determinations of
the positions with respect to a fundamental reference source were
made. This then agrees well with the estimate of $\approx 3$ mas
accuracy for the {\it absolute} position of the calibrators. From the
scatter in the positions of the calibrators with respect to each
other, we conclude that the errors in the {\it relative} astrometry,
due to the phase-referencing, are of the order of $1.6$ mas in each
coordinate for each epoch.

The positions of several maser spots were measured at each epoch and
used to determine their motions. One should bear in mind that the
image quality differs substantially between epochs; not only is the
quality of the data dependent on ionospheric conditions (solar cycle,
day and night) and elevation, but also on the OH variability of U~Her.
For our astrometric purposes, the images have been processed without
using self-calibration so that the phase information is preserved. The
final spectral resolution used was 2 kHz (Fig.\ \ref{motion}).

\section{Results}

\subsection{Maser spots}

On VLBI baselines most ($\approx 75\% $) of the OH maser emission is
resolved out. Although the U~Her 1667~MHz line displays a regular
double-peaked spectrum in single dish observations, the VLBI emission
is limited to blue-shifted emission only. Maser spots are found in a
velocity range of $-17.5$ to $-20.6$ km/s, the stellar velocity of
U~Her is $v_{\rm lsr} = -14.5$ km/s. In different epochs 4 -- 7
regions of VLBI emission can be found, spread over $\approx$ 200 mas.
The brightest maser spots are resolved and have sizes between 14 and
25 mas, even in the few cases where we can self-calibrate the
data. Moreover, structure more compact than $\approx 20$ mas has
seldom been detected previously in circumstellar OH main line
masers. Therefore we believe that the fact that the maser spots are
resolved is not an artifact of the phase referencing.

\begin{figure}
   \resizebox{\hsize}{!}{\rotatebox{-90}{\includegraphics{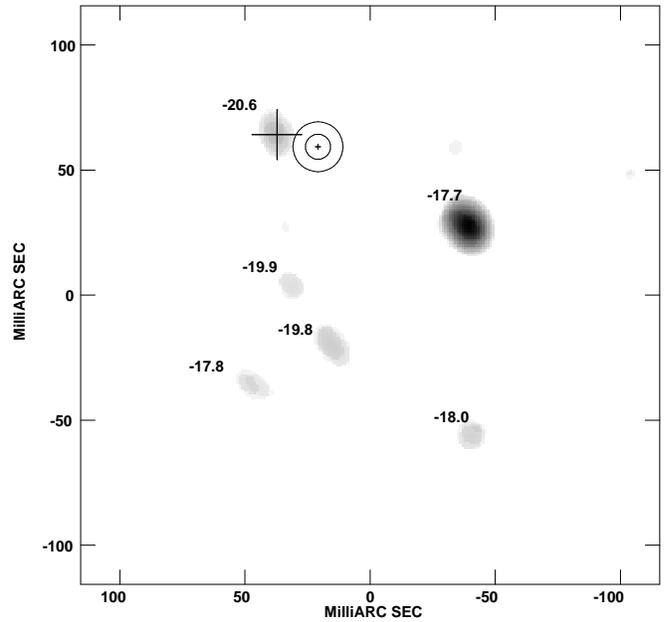}}}
   \hfill
   \caption{ A zeroth moment map of the maser spots at the first
   observational epoch. Each spot is labeled with the corresponding
   velocity. The optical and radio photosphere of the star are drawn
   (two concentric circles) as well as the error bars on optical and
   radio positions. The error bars on the radio position include the
   uncertainty from extrapolating the fitted proper motion. Note that
   for this epoch the most blue-shifted spot was not the brightest.}
   \label{mom}
\end{figure}

The most blue-shifted maser spot can be identified at all epochs
(Fig.\ \ref{motion}).  However it does not seem to be any brighter
(stronger or more compact) than some of the other compact spots
(Fig.~\ref{mom}) (although one should bear in mind that the phase
referencing does not calibrate the phases perfectly). Most of the
other spots can be tracked over many epochs as well.

Although the 1665 MHz OH maser was observed with only modest spectral
resolution it has been possible to trace the maser spots for
all 6 epochs. The low resolution prohibited accurate determination of
the velocities of the individual maser spots. However, it was still
possible to pair many 1667 and 1665 MHz spots, solely on
the basis of their relative positions.

Approximately 75\% of the brightest spots detected at 1665 MHz
coincide with 1667 MHz spots. Typically the positions match to
$\approx 6$ mas which is slightly less than the beam size and
certainly smaller than the maser spot size. An accurate determination
of the flux ratio is difficult because of the difference in spectral
resolution but the 1665 MHz maser spots generally appear to be the
brightest. The trend of decreasing flux ratio with decreasing velocity
(Sivagnanam et al.\ \cite{SivaAmpl}) is not supported by these
observations.

\subsection{Amplified stellar image}

Observations of U~Her and other OH-masering stars have revealed compact
maser sources on the blue side of the circumstellar shell, but no
corresponding red-shifted features (Sivagnanam et
al. \cite{SivaAmpl}). This has been suggested to be due to
amplification of continuum radiation from the stellar radio-sphere; the
amplified stellar image theory. By comparing the optical stellar
position of U~Her as determined by Hipparcos with the positions of the
compact maser features we have tested this hypothesis.

Previous attempts to compare radio and optical positions for maser
stars had to rely on ground-based optical data (Baudry et al.\
\cite{Baud}). Comparisons have been made using OH positions (Bowers
\& Johnston\ \cite{Bow2}), H$_{2}$O positions (Bowers et al.\
\cite{Bow1}) and SiO positions (Wright et al.\ \cite{Wrig}; Baudry et
al.\ \cite{Baud2}). These studies were limited to $\approx 60$ mas
accuracy. With the current measurements and the data from the
Hipparcos mission available, the accuracy can be increased
considerably. Positions and proper motions in the Hipparcos catalogue
are given for the reference epoch J1991.25 within the International
Celestial Reference System (ICRS). The alignment of the Hipparcos
frame with the ICRS was made at $\la$ 1 mas accuracy through several
link programs among which VLBI observations of radio stars played a
central role (Lestrade et al.\ \cite{Lestrade}). As noted before, our
reference sources are tied to the fundamental reference frame using
the USNO frame with an accuracy of $\la$ 3 mas.

Comparison of the radio and optical position requires a transposition
of the radio position to the mean epoch of the Hipparcos
observations, J1991.25. For this we use the proper motion and parallax
as determined by our fits (see below). The error in the transposed
radio position is therefore dominated by the error in proper motion.
For our first observational epoch this results in an error of $\approx
3$ mas, for the last epoch the error has increased to $\approx 7$ mas.
Taking into account the error in the parallax, the error in the
transposed position can be estimated to be 5 -- 9 mas in each
coordinate. Combined with the error in the radio position
determination and the Hipparcos position error this allows us to
compare the optical and radio positions with $\approx$ 15 mas accuracy
(Fig.~\ref{mom}).

\begin{figure}
   \resizebox{8cm}{!}{\includegraphics{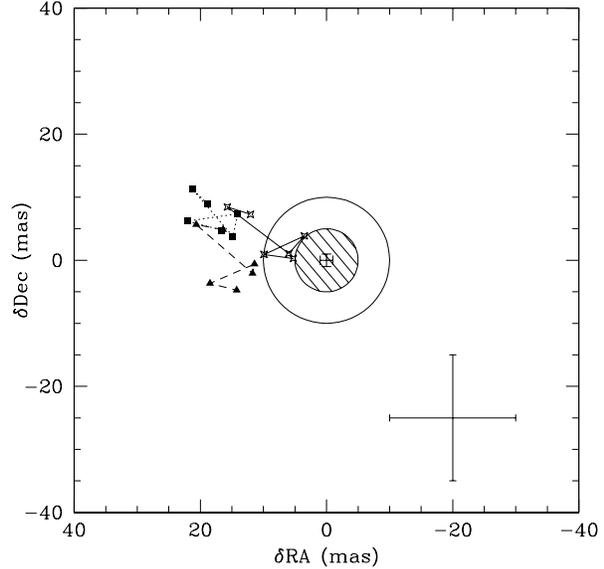}}
   \hfill
   \caption{The position of the most blue-shifted maser spot wrt the
   optical position as predicted by our fits. Points of subsequent
   epochs are connected. The squares correspond to the fit to the most
   blue-shifted spot only, the triangles correspond to a fit to the
   three brightest spots, the stars are the fit to 1 spot using the
   Hipparcos position as zero-point. Also drawn are indications of the
   size of the star (shaded) and the radio-photosphere. The error bars
   in the bottom right corner are the errors on the transposed radio
   positions, drawn for the first epoch. Subsequent epochs have
   slightly larger error bars. The error bars on the star are the
   Hipparcos position errors.}  \label{scat}
\end{figure}

Fig.~\ref{scat} shows the deviation between the radio observations and
the predicted optical position using different fits for the proper
motion. The radio photosphere of U~Her can be estimated by using the
SiO maser observations by Diamond \& Kemball\ \cite{Diam2}. They
provide an upper limit of $\approx 20$ mas if, as proposed by Reid \&
Menten\ \cite{Reid}, the radio photosphere extends to the edge of the
SiO masering region. The size of the optical photosphere of the star
is thought to be half that of the radio photosphere, thus the stellar
disk is expected to have a diameter of $\approx 10$ mas. This value
is consistent with a straightforward diameter estimate, assuming
fundamental mode pulsation and a distance of 189 pc. As seen in
Fig.~\ref{scat} the deviations are not significantly different for the
various fits that were performed. There seems to be a systematic shift
of $\approx 15$ mas in right ascension.
 
If the most blue-shifted spot originates from amplification of a
smooth background source by a masering screen, one may expect the size
of the maser spot to be comparable to that of the stellar
radio-sphere. The spot is found to have a diameter of $20.4 \pm 3.4$ mas,
consistent with the estimated radio-sphere size of 20 mas. The
centroid could wander up to $\approx 6$ mas due to temporal variations of
the maser or the background stellar continuum.  Indeed we detect some
position variations in Fig.~\ref{scat}, which seem to be of the order
of the assumed diameter of the stellar photosphere, as expected

\begin{figure*} 
   \resizebox{12cm}{!}{\includegraphics{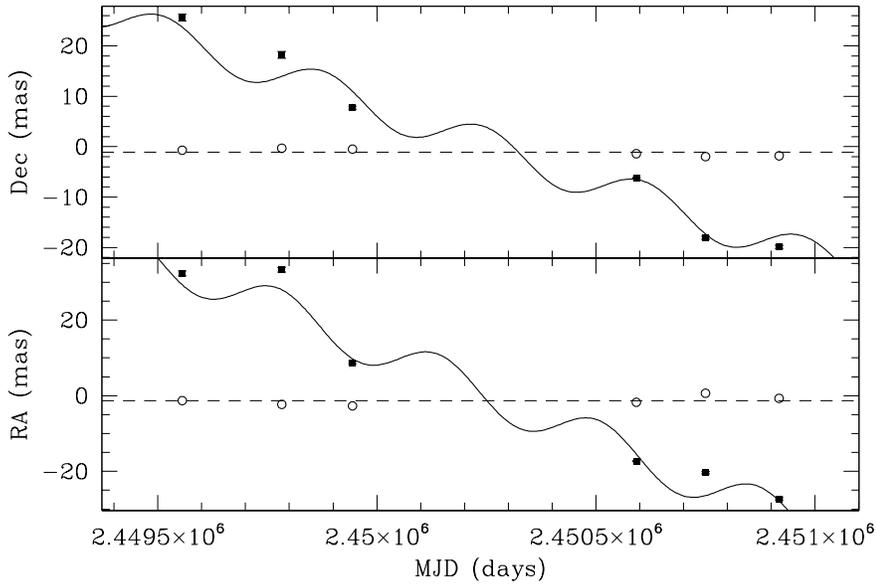}}
   \hfill \parbox[b]{55mm} { \caption{The difference between the
   measured separation and its a-priori value of 1636+2112 (open
   symbols) and of the most blue-shifted 1667~MHz maser
   spots (filled symbols) with respect to 1628+214. The formal error
   bars on the position of 1636+2112 are smaller than the symbol
   size. Drawn are the measured average separation for the two
   extra-galactic sources (dashed) and the best fitting parallax and
   proper motion trajectory for U~Her (solid) when using the most blue-shifted
   spot.}  \label{parlx} }
\end{figure*} 

\subsection{Parallax \& proper motion}

The proper motion and parallax were determined using a least-square
fitting method on the 1667 MHz data. A fit to the motion of the most
blue-shifted maser spot, visible in all 6 epochs, resulted in
$\mu_{\rm VLBI} = -17.4 \pm 1.5$, $-10.9 \pm 1.4$ mas/yr for the
proper motion in right ascension and declination
respectively (throughout the paper the errors represent $1 \sigma$
deviations). If we include two other bright spots, each of which we
could detect at 5 of the 6 epochs, we find: $\mu_{\rm VLBI} = -17.05
\pm 0.85$, $-9.48 \pm 0.73$ mas/yr.  Both proper motion values agree
well with the result on U~Her by Hipparcos: $\mu_{\rm Hip} = -16.84
\pm 0.82, -9.83 \pm 0.92$ mas/yr.  The Hipparcos observations (ESA
\cite{ESA}) give highly accurate positions but a rather poor
trigonometric parallax ($\pi_{\rm Hip} = 1.64 \pm 1.31$ mas), so the
distance to the star is not reliably known. Numerical analysis shows
that the VLBI data is significantly better represented by a model
which includes a parallax.  The best fit using the VLBA data for only
the brightest spot gives $\pi_{\rm VLBI} = 5.3 \pm 2.1$ mas
(Fig.~\ref{parlx}). The fit to three spots simultaneously gives
$\pi_{\rm VLBI} = 4.2 \pm 1.2$ mas.  Although the fit to the three
brightest maser spots results in smaller errors on the parallax and
proper motion this result has to be treated with care. There is no
clear reason to assume that all masers spots remain fixed with respect
to each other, in fact they do not (see below).  Only the most
blue-shifted spot is expected to remain fixed to the star. Using this
we have also performed a fit on that spot including the Hipparcos
optical position as additional data point.  This resulted in $\mu =
-15.8 \pm 1.0$, $-10.20 \pm 1.0$ mas/yr and $\pi = 4.1 \pm 2.0$ mas.
However, due to the difficulties in the connection of the optical and
radio frames systematic errors may be present. We adopt the fit of
$\pi_{\rm VLBI} = 5.3 \pm 2.1$ mas and $\mu_{\rm VLBI} = -17.4 \pm
1.5$, $-10.9 \pm 1.4$ mas/yr, with a single maser spot and radio data
only as our most reliable estimate of the parallax and proper motion.

The errors displayed in Fig.\ \ref{parlx} are the formal uncertainties
in fitting Gaussian profiles to the maser spot in order to obtain the
position. These errors are typically $\approx 1.0$ mas in each
coordinate, a fraction of the beam. The additional systematic errors
in the {\it relative} astrometry originating from the phase referencing were
estimated to be $\approx 1.6$ mas.  The final fit, however, leaves
rms residuals of $3.7$ and $2.5$ mas in right ascension and
declination respectively. The maximum deviation is $6.1$ mas in right
ascension and $4.3$ mas in declination. 

As noted before, the separation between the bright maser spots is not
constant. Between the different observational epochs shifts average
$\approx$ 2.5 mas. The separation between the most-blue-shifted spot
and the spot at -17.7 km/s, which is the brightest spot at the first
epoch, seems to show some systematic expansion. However, it is possible
that between the second and third epoch the spot at -17.7 km/s
disappears and thus we might be tracking a different spot afterwards.

 The residuals and the relative spot motion can be only partly
attributed to turbulent motion in the masering shell. A typical value
of 1 km/s turbulent motion (Diamond et al.\ \cite{Diam}) corresponds
to 0.7 mas/yr at a distance of 300 pc.

\section{Discussion}

One motivation to attempt the detection of the U~Her motion by means
of the OH maser, was the prediction that on the line of
sight to the star the maser will amplify radio photons from the
stellar surface (Norris et al.\ \cite{Norris}; Van~Langevelde \&
Spaans \cite{Spaans2}). In a spherically symmetric shell with a
constant outflowing velocity, the most blue-shifted part of the
masering shell would then be amplifying a background source rather
than spontaneous emission. Hence this would result in a spot with more
maser beaming and thus brighter, more compact emission. The spot would
be fixed to the stellar radio-photosphere (which could be as large as
$\approx 20$ mas for U~Her, while the visible photosphere is perhaps
twice smaller) and probably be persistent over the
years. It would also be observable in different maser lines, as the
mechanism would work in a similar way for all transitions. It could still
vary in flux with stellar cycle through changing pumping conditions.

This idea is confirmed by our results, although some issues remain
unexplained. We find that the brightest and most compact features are
all from the blue-shifted side of the shell. We find no high
brightness red-shifted maser spots. Our wide-band data confirm the
observation by Sivagnanam et al.\ (\cite{SivaAmpl}) that the most
blue-shifted 1667~MHz spot matches one of the 1665 MHz spots
exactly. This spot at $-20.6$ km/s also seems the most persistent. We
managed to follow this 1667~MHz and corresponding 1665~MHz features
for over 4 years.  The size of the most blue-shifted spot is
consistent with the estimated size of the stellar radio-sphere.
Finally, and most importantly, we have been able to compare the radio
and optical positions. The position of the most blue-shifted maser
spot has been shown to match the Hipparcos optical position within the
errors.

However, contrary to what would be expected, the most blue-shifted
spot is not always the brightest or most compact; other spots at
different velocities are dominant at different epochs. And not just
the most blue-shifted 1667~MHz maser spot coincides with a 1665~MHz
spot; almost $75\%$ of the other bright spots coincide with one as
well. Furthermore, it should have been possible to describe the motion
of the $-20.6$ km/s spot, presumably fixed on the stellar position,
with a simple trajectory. So, although our data is consistent with the
amplified stellar image theory, the residuals of the fit may be
indicative of more complex kinematics.  A fixed maser spot could still
show some random motion, as it drifts over the large surface of the
radio-photosphere. The compact maser spot is a result of high beaming
so small motions could result from minor changes of the maser path
length due to variations in the pumping mechanism. This could also
explain some of the relative motion between the spots. A correlation
between stellar phase and these motions however cannot be
determined. One other possibility is that U~Her itself does not follow
a simple trajectory because it is part of a binary or multiple
system. All these mechanisms could give rise to residuals of the order
of those observed. ($3.7$, $2.5$ mas rms)

Another issue that remains is the high surface brightness of the
multiple maser spots. The most blue-shifted spot can be explained by
amplified emission from the stellar radio-photosphere, but it is not
always the brightest spot. The brightness of these other spots cannot
easily be explained by normal self-amplifying masering regions. Since
they are also only seen at the blue-shifted side of the shell, and
since they move along with the star, they can not be amplified
sources from beyond the shell. They lie too far from the star to
be amplifying part of a very extended stellar radio-photosphere. At least
two other spots are too bright to be self-amplifying. Should U~Her
really be a multiple system, the high brightness spots could be the
other components, amplified like U~Her itself. The motion of a
multiple system should also be visible in the separation of the spots
at different epochs. There might be an indication of such systematic
motion between two of the spots. The occurrence of multiple bright
spots is a subject of further study in a larger sample of OH-masering
stars and will also be studied in the 1612 MHz maser transition.

Although the precise nature of the additional compact blue-shifted
emission features remains debatable, amplified stellar emission seems
indeed to be the cause of the compact most blue-shifted maser spot.
We are able to measure a proper motion derived from the VLBI maser
spot which is consistent with the results from the Hipparcos
satellite. The fit to the observed trajectories improves significantly
when a parallax is included. The distance of $189 (+123, -54)$ pc
determined with VLBI from a fit to the most blue-shifted OH spot is
significantly smaller than the Chapman et al.\ \cite{Chapman} $P-L$
estimate of 385 pc. It is somewhat smaller than the revised $P-L$
estimate of 280 pc by Alvarez \& Mennesier (\cite{AlMe}).

{\it Acknowledgments:} 
We thank Craig Walker for help verifying the correlator model used,
Jim Brauher for carrying out some initial data reduction and Jean-Francois
Lestrade and Bob Campbell for providing software to check our fitting
procedures. This project is supported by NWO grant 614-21-007.


\begin{thebibliography}{999}




\bibitem[1997]{AlMe}
Alvarez, R., Mennesier, M.-O., 1997, AA, 317, 761.

\bibitem[1990]{Baud}
Baudry, A., Mazurier, J.M., Peri\'e, J.P., Requi\`eme, Y., Rousseau,
J.M., 1990, AA, 232, 258.

\bibitem[1995]{Baud2}
Baudry, A., Lucas, R., Guilloteau, S., 1995, AA, 293, 594.

\bibitem[1994]{Bow2}
Bowers, P.F., Johnston, K.J., 1994, ApJSS, 92, 189.

\bibitem[1989]{Bow1}
Bowers, P.F., Johnston, K.J., de Vegt, C., 1989, ApJ, 340, 479.

\bibitem[1994]{Chapman}
Chapman, J.M, Sivagnanam, P., Cohen, R.J., Le Squeren, A.M, 1994, 
MNRAS, 268, 475.

\bibitem[1994]{Diam2}
Diamond, P.J., Kemball, A.J., Junor, W., Zensus, A., Benson, J.,
Dhawan, V., 1994, ApJ, 430, L61.

\bibitem[1985]{Diam}
Diamond, P.J., Norris, R.P., Rowland, P.R., Booth, R.S.,
    Nyman, L.-\AA., 1985, MNRAS, 212, 1.

\bibitem[1997]{ESA}
ESA, 1997, {\it The Hipparcos and Tycho Catalogues}, ESA SP-1200.

\bibitem[1995]{Lestrade}
Lestrade, J.-F., Jones, D.L., Preston, R.A. et al.,\ 1995, AA, 304, 182.

\bibitem[1990]{Ma}
Ma, C., Shaffer, D.B., De~Vegt, C., Johnston, K.J., Russell, J.L., 1990, AJ,
99, 1284.

\bibitem[1984]{Norris}
Norris, R.P., Booth, R.S., Dia\-mond, P.J., Ny\-man, L.A., Gra\-ham, D.A., 
Matveyenko, L.I., 1984, MNRAS, 208, 435.

\bibitem[1992]{Patnaik}
Patnaik, A.R., Browne, I.W.A., Wilkinson, P.N., Wrobel, J.M., 
1992, MNRAS, 254, 655.

\bibitem[1997]{Reid}
Reid, M.J., Menten, K.M., 1997, ApJ, 476, 327.

\bibitem[1990]{SivaAmpl}
Sivagnanam, P., Diamond, P.J., Le Squeren, A.M., Biraud, F., 
1990, AA, 229, 171.

\bibitem[1993]{Spaans2}
Van~Langevelde, H.J., Spaans, M., 1993, MNRAS, 264, 597. 

\bibitem[1998]{Wilkinson}
Wilkinson, P.N., Browne, I.W.A., Patnaik, A.R., Wrobel, J.M., Sorathia, B.,
1998, MNRAS, 300, 790.

\bibitem[1990]{Wrig}
Wright, M.C.H., Carlstrom, J.E., Plambeck, R.L., Welch, W.J., 1990,
AJ, 99, 1299.

\bibitem[1994]{YatCoh}
Yates, J.A., Cohen, R.J., 1994, MNRAS, 270, 958.

\end{thebibliography}
\end{document}